\documentclass{jfm}

\usepackage{graphicx}
\usepackage{natbib}

\usepackage{graphicx}
\usepackage{amsmath}
\usepackage{amssymb}

\newcommand{\be}{\begin{equation}}
\newcommand{\ee}{\end{equation}}
\newcommand{\bs}{\begin{subequations}}
\newcommand{\es}{\end{subequations}}
\newcommand{\rmd}{\mathrm{d}}
\newcommand{\rme}{\mathrm{e}}
\newcommand{\rmi}{\mathrm{i}}

\newcommand{\efac}{\rme^{\rmi\mathbf{k}\cdot{\boldsymbol \xi}}}
\newcommand{\Fr}{\mathrm{Fr}}
\newcommand{\Frs}{\mathrm{Fr}_\mathrm{s}}

\newcommand\eg{e.g.\ }

\ifCUPmtlplainloaded \else
  \checkfont{eurm10}
  \iffontfound
    \IfFileExists{upmath.sty}
      {\typeout{^^JFound AMS Euler Roman fonts on the system,
                   using the 'upmath' package.^^J}%
       \usepackage{upmath}}
      {\typeout{^^JFound AMS Euler Roman fonts on the system, but you
                   dont seem to have the}%
       \typeout{'upmath' package installed. JFM.cls can take advantage
                 of these fonts,^^Jif you use 'upmath' package.^^J}%
       \providecommand\upi{\upi}%
      }
  \else
    \providecommand\upi{\upi}%
  \fi
\fi

\ifCUPmtlplainloaded \else
  \checkfont{msam10}
  \iffontfound
    \IfFileExists{amssymb.sty}
      {\typeout{^^JFound AMS Symbol fonts on the system, using the
                'amssymb' package.^^J}%
       \usepackage{amssymb}%
         \let\leq=\leqslant
         \let\geq=\geqslant
      }{}
  \fi
\fi

\ifCUPmtlplainloaded \else
  \IfFileExists{amsbsy.sty}
    {\typeout{^^JFound the 'amsbsy' package on the system, using it.^^J}%
     \usepackage{amsbsy}}
    {\providecommand\boldsymbol[1]{\mbox{\boldmath $##1$}}}
\fi

\title[Ship waves with vorticity]{Ship waves in the presence of uniform vorticity}
\author[S. \AA. Ellingsen]{Simen \AA. Ellingsen$^1$\thanks{Email address for correspondence: simen.a.ellingsen@ntnu.no}}

\affiliation{$^1$Department of Energy and Process Engineering, Norwegian University of Science and Technology, N-7491 Trondheim, Norway}

\begin{document}

\maketitle
\begin{abstract}
  Lord Kelvin's result that waves behind a ship lie within a half-angle $\phi_\mathrm{K}\approx 19^\circ28'$ is perhaps the most famous and striking result in the field of surface waves. We solve the linear ship wave problem in the presence of a shear current of constant vorticity $S$, and show that the Kelvin angles (one each side of wake) as well as other aspects of the wake depend closely on the ``shear Froude number'' $\Frs=VS/g$ (based on length $g/S^2$ and the ship's speed $V$), and on the angle between current and the ship's line of motion. In all directions except exactly along the shear flow there exists a critical value of $\Frs$ beyond which no transverse waves are produced, and where the full wake angle reaches $180^\circ$. Such critical behaviour is previously known from waves at finite depth. For side-on shear, one Kelvin angle can exceed $90^\circ$. On the other hand, the angle of maximum wave amplitude scales as $\Fr^{-1}$ ($\Fr$ based on size of ship) when $\Fr\gg1$, a scaling virtually unaffected by the shear flow.
\end{abstract}

\section{Introduction}

The striking V-shaped wave pattern produced by a boat or swimming duck is familiar to all. A beautiful sight from a lakeside, it is also of pivotal importance in marine technology.
Shown by Lord Kelvin in 1887 \citep{kelvin1887}, the result that the train of waves in a ship's wake lie within a half-angle of $\phi_\mathrm{K}=\arcsin(1/3)\approx 19^\circ28'$ is one of the most famous and celebrated in the field of surface waves. Remarkably, although qualitative aspects of the ship waves depend on physical parameters, the Kelvin angle does not --- under the assumptions of deep water and no shear flow, it is universal. 

Recently, an investigation of ship waves by aerial photography by \citet{rabaud13} indicated that when the Froude number $\Fr=V/\sqrt{gb}$ ($b$ is the size of the wave source and $V$ its speed) becomes high, the wake angle decreased as $\Fr^{-1}$, apparently running counter to Kelvin's result. This was confirmed theoretically by \citet{darmon13} by noting that while Kelvin's derivation is still valid, the angle of maximum wave amplitude scales as $\Fr^{-1}$, similar to the Mach angle of a shock-wave. 

Typically, more than 30\% of a ship's energy consumption is due to wave resistance, energy propagated away via the ship's wake \citep{faltinsen05}, so understanding ship waves in various environments is highly technologically important. 
There are several common situations in which shear flow is of importance to surface waves \citep{peregrine76}, including wind induced drift of the surface layer of water, shallow flow over a sea bed, and in the presence of underwater currents. A model system where vorticity is assumed constant has been the a standard study \citep[\eg][]{taylor55}, combining nontrivial physics with mathematical tractability. While only an approximation of real situations, the constant vorticity model has been found to reproduce essentials of observations (\eg\cite{evans55}). 
We show in this article that when a constant vorticity $S$ is present below the ship (hereafter called the wave source), Kelvin's angle depends on a second Froude number, 
\be
  \Frs = VS/g
\ee
based on the shear length $l_S=g/S^2$. Physically, $z=-2l_S$ is the depth at which potential and kinetic energy densities of the shear flow are equal (factor $2$ dropped for simplicity).

Following Kelvin's seminal work, the waves behind a moving pressure source were investigated by several authors, \cite{havelock08,havelock19} perhaps most prominent among them --- a thorough review of the classical literature is \citet[Sec.~13]{wehausen60}. A more recent review is \cite{reed02}.
Naturally, numerous examinations of the interaction between waves and shear currents exist \citep[see {\eg}][]{telesdasilva88, peregrine76,brevik76,fabrikant98,buhler09}, yet to our knowledge neither concerns how the Kelvin wake is affected by shear, although a similar problem has recently been considered by \cite{benzaquen12} in two dimensions with particular focus on wave resistance, and for the wake behind a partly submerged cylinder, in two dimensions, by \cite{mccue99}. Closest is perhaps \cite{peregrine71} who considered the interaction between ship waves and the ship's own viscosity-induced wake.  We shall not consider this effect in the present endeavour.

\section{Formalism and solution of linear problem}

\begin{figure}
\begin{center}
  \includegraphics[width=.6\columnwidth]{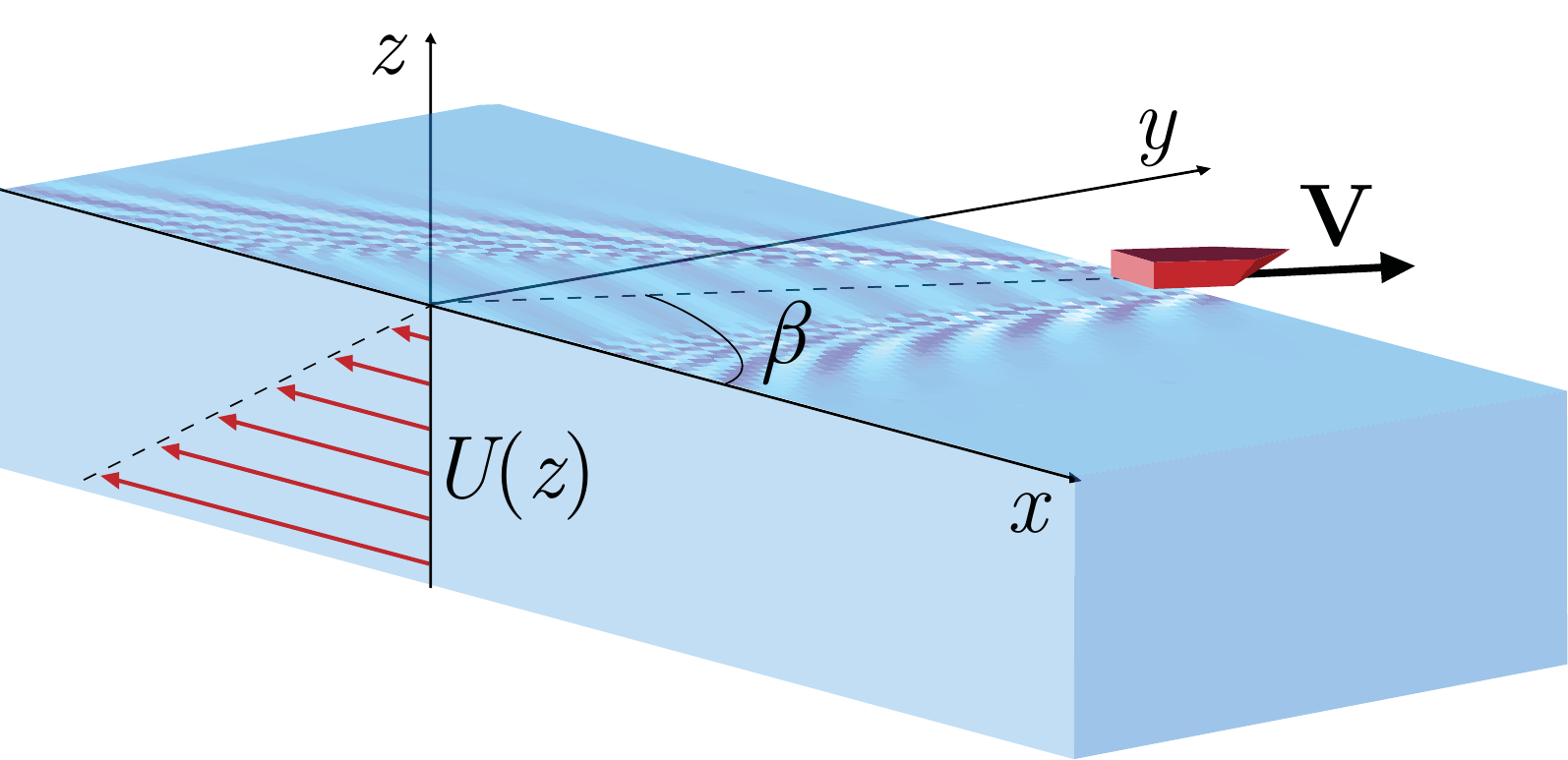}
\end{center}
\caption{The geometry: a Couette profile current interacts with surface waves generated by a source moving at constant velocity $\mathbf{V}$.}
\label{fig:geom}
\end{figure}

Consider the system depicted in figure \ref{fig:geom} in which the basic flow --- a Couette flow profile --- is disturbed in such a way as to produce waves on the surface of the liquid. To keep the number of parameters down we assume the liquid has infinite depth and negligible surface tension (generalisation is straightforward). This flow is essentially three dimensional and rotational, hence potential theory is not an option \citep{ellingsen13}. We assume the liquid is incompressible and of density $\rho$.

We write the full velocity field as
\be
  \mathbf{v} = (U(z)+\hat{u},\hat{v},\hat{w});~~ U(z) = Sz,
\ee
and $S\geq 0$ is the uniform vorticity. $U(z)$ is the basic shear flow, and $\hat{u},\hat{v},\hat{w}$ small velocity perturbations. When the vorticity is uniform no critical layers can form beneath the surface, see \citet{booker67} and \citet[ch.~7]{leblond78}. We have assumed the surface of water to be at rest with respect to our coordinate system --- this is easily generalised by an overall Galilean transformation. 

The situation considered is one in which a wave source in the form of a pressure distribution travels at a constant velocity $\mathbf{V}=(V\cos\beta,V \sin\beta)$ in the $xy$ plane. We consider only stationary solutions as seen by the source, so that all physical quantities depend on $\mathbf{x}=(x,y)$ and time $t$ only through the combination ${\boldsymbol\xi} = \mathbf{x} - \mathbf{V}t$.

We presume a periodic perturbation in the ${\boldsymbol\xi}$ plane:
\be
  [\hat{u},\hat{v},\hat{w}] = [u(z),v(z),w(z)]\efac
\ee
with wave vector $\mathbf{k}=(k_x,k_y)= (k\cos\theta,k\sin\theta)$.
Also the pressure is perturbed from the basic distribution (the hydrostatic pressure),
\be
  P({\boldsymbol\xi},z) = \hat{p}- \rho g z = p(z)\efac - \rho g z.
\ee

The waves are produced by a constant external pressure source $p_\text{ext}({\boldsymbol \xi})$ moving along the surface at velocity $\mathbf{V}$, whose Fourier transform $\tilde{p}(\mathbf{k})$ is defined by
\be
  p_\text{ext}({\boldsymbol \xi}) = \int \frac{\rmd^2 k}{(2\upi)^2}\tilde{p}(\mathbf{k})\efac.
\ee
Likewise the observable values of other physical quantities are real parts of their integrals over all values of $\mathbf{k}$: $\hat{u}({\boldsymbol\xi},z) = \int \rmd^2 k(2\upi)^{-2}u(z)\efac,$
etc.

We work to linear order in the small quantities $u(z), v(z), w(z)$ and  $p(z)$ and we shall assume that viscous effects may be neglected. Inserting into the Euler equation and continuity equation $\nabla\cdot\mathbf{v}=0$ then gives after linearisation:
\begin{align}
  \rmi k_x u + \rmi k_y v + w' =& 0; & 
  \rmi(k_xU-\mathbf{k}\cdot\mathbf{V})u + S w =& -\rmi k_xp/\rho;\notag\\
  \rmi(k_xU-\mathbf{k}\cdot\mathbf{V})v =& -\rmi k_yp/\rho; &
  \rmi(k_xU-\mathbf{k}\cdot\mathbf{V})w =& -p'/\rho.\label{Ec}
\end{align}

Now eliminating 
$u,v$ and $p$ 
we obtain $
  w'' = k^2 w$ in accordance with the Rayleigh equation \citep[see][p.~390]{leblond78}, 
which, when subjected to the boundary condition that $w(-\infty)=0$, yields a general solution to \eqref{Ec},
\begin{align}
  u =& \rmi A(\mathbf{k}) \rme^{kz}\Bigl[k_x + \frac{Sk_y^2}{k(k_xU-\mathbf{k}\cdot\mathbf{V})}\Bigr], &
  v =& \rmi A(\mathbf{k}) \rme^{kz}\Bigl[k_y - \frac{Sk_xk_y}{k(k_xU-\mathbf{k}\cdot\mathbf{V})}\Bigr], \notag\\
  w =& kA(\mathbf{k}) \rme^{kz},  &
  \frac{p}{\rho} =& -\rmi A(\mathbf{k})\rme^{kz}\Bigl(k_xU-\mathbf{k}\cdot\mathbf{V}-\frac{Sk_x}{k}\Bigr);\label{sol}
\end{align}
where $A(\mathbf{k})$ is spatially and temporally constant.

We define the surface elevation (relative to equilibrium)
\be\label{zB}
  \zeta({\boldsymbol\xi}) = B(\mathbf{k}) \efac.
\ee
The linearised kinematic boundary condition at the surface is $\partial_t \zeta = \hat{w}|_{z=0}$, and the dynamic boundary condition, when surface tension is ignored, is that $P({\boldsymbol\xi},\zeta) = p_\text{ext}$. Inserting Eqs.~\eqref{sol} and \eqref{zB} this implies
\be
  -\rmi(\mathbf{k}\cdot\mathbf{V}) B(\mathbf{k}) = kA(\mathbf{k}), ~~ \text{and} ~~ -\rmi A(\mathbf{k})[\mathbf{k}\cdot\mathbf{V} +S(k_x/k)]-gB(\mathbf{k})= \tilde{p}(\mathbf{k})/\rho.
\ee
Combining these and eliminating $A(\mathbf{k})$ yields
\be\label{B}
  B(\mathbf{k}) = -\frac1{\rho}\frac{k\tilde{p}(\mathbf{k})}{gk-(\mathbf{k}\cdot\mathbf{V})^2 -S(\mathbf{k}\cdot\mathbf{V})\cos\theta}.
\ee

Applying the procedure of \cite{ellingsen13} with the present formalism, one readily shows that the phase velocity of a wave propagating in direction $\mathbf{k}$ is unique, positive, and equals\footnote{
At finite depth $h$ and including surface tension coefficient $\sigma$ this generalises to $c(\mathbf{k}) = [c_0^2(k)\tanh kh+(S/2k)^2\cos^2\theta\tanh^2 kh]^{1/2}-(S/2k)\cos\theta\tanh kh$ with $c_0^2=g/k+k\sigma/\rho$.}
\be  \label{deep}
  c(\mathbf{k}) = \sqrt{\frac{g}{k}+\Bigl(\frac{S}{2k}\cos\theta\Bigr)^2}-\frac{S}{2k}\cos\theta.
\ee
Only waves $\mathbf{k}\cdot\mathbf{V}>0$ can propagate along with the boat and give stationary solutions, so angles $\theta$ between $\beta-\upi/2$ and $\beta+\upi/2$ need be integrated. Let moreover ${\boldsymbol\xi}=(r\cos\phi,r\sin\phi)$. Then, with $B(\mathbf{k})$ as given in Eq.~\eqref{B},
\be
  \zeta({\boldsymbol\xi}) = \int_0^\infty\frac{\rmd k}{2\upi}\int_{\beta-\upi/2}^{\beta+\upi/2}\frac{\rmd\theta}{2\upi}kB(\mathbf{k})\rme^{\rmi kr\cos(\theta-\phi)}.\label{zetaint}
\ee
In the limit $S\to0$ and $\beta=0$, the expression due to \cite{havelock19} is regained. 

We choose the same Gaussian pressure source as used by \cite{darmon13},
\be
  p_\text{ext}({\boldsymbol\xi})=p_0 \rme^{-\upi^2\xi^2/b^2}; ~~ \tilde{p}(\mathbf{k})=(b^2p_0/\upi)\rme^{-k^2b^2/(2\upi)^2}
\ee
where $b$ is the characteristic length of the pressure source and basis for the Froude number $\Fr=V/\sqrt{gb}$.

\begin{figure}
\begin{center}
  \includegraphics[width=\columnwidth]{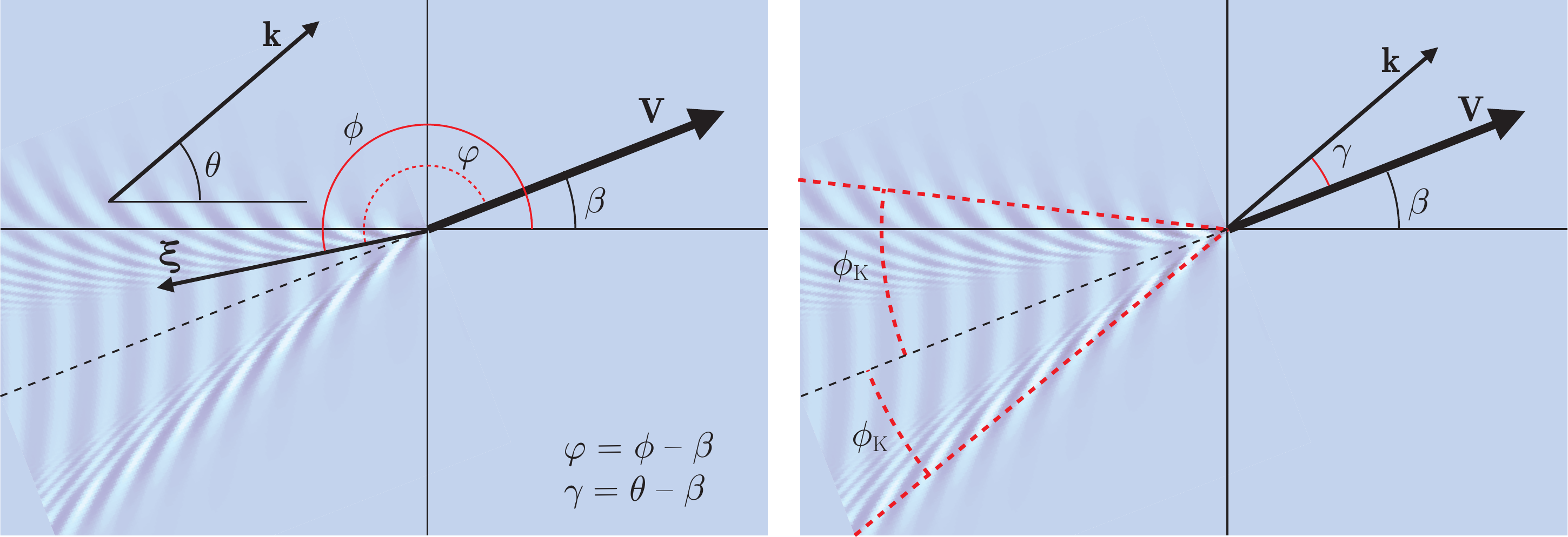}
\end{center}
\caption{Definition of angles used in the article. Polar angles are $\phi$ and $\theta$ in the ${\boldsymbol\xi}$ and $\mathbf{k}$ planes, respectively.}
\label{fig:angles}
\end{figure}

We proceed like \cite{darmon13} to make integral \eqref{zetaint} definite by means of the radiation condition. Letting all perturbation quantities $\propto \rme^{\tilde\epsilon t}$ for a positive infinitesimal $\tilde\epsilon$ is mathematically equivalent to adding a small imaginary part to the velocity; $\mathbf{k}\cdot\mathbf{V}\to \mathbf{k}\cdot\mathbf{V}+\rmi\epsilon$, taking $\epsilon\to 0$ in the end. The integral over $k$ can then be evaluated using the Sokhotsky-Plemelj theorem \citep[see][]{darmon13,raphael96},
\be\label{sp}
  \lim_{\epsilon\to 0}\int_a^b \frac{f(x)\rmd x}{x\pm \rmi\epsilon} = \mp \rmi\upi f(0) + \mathcal{P}\int_a^b \frac{f(x)}{x}
\ee
where $\mathcal{P}$ is principal value. The last term of Eq.~\eqref{sp} may be shown to vanish quickly as $r\to \infty$ \citep{raphael96}; it is the surface depression near the source and is discussed in detail \eg by \citet[Sec.~13]{wehausen60}. It depends closely on the specific shape of $p_\text{ext}$, and since it does not describe propagation towards infinity, it cannot contribute to the transportation of energy away from the source, hence it does not add to the wave resistance. We shall be interested only in the first term.

\begin{figure}
  \begin{center}
  \includegraphics[width=.32\textwidth]{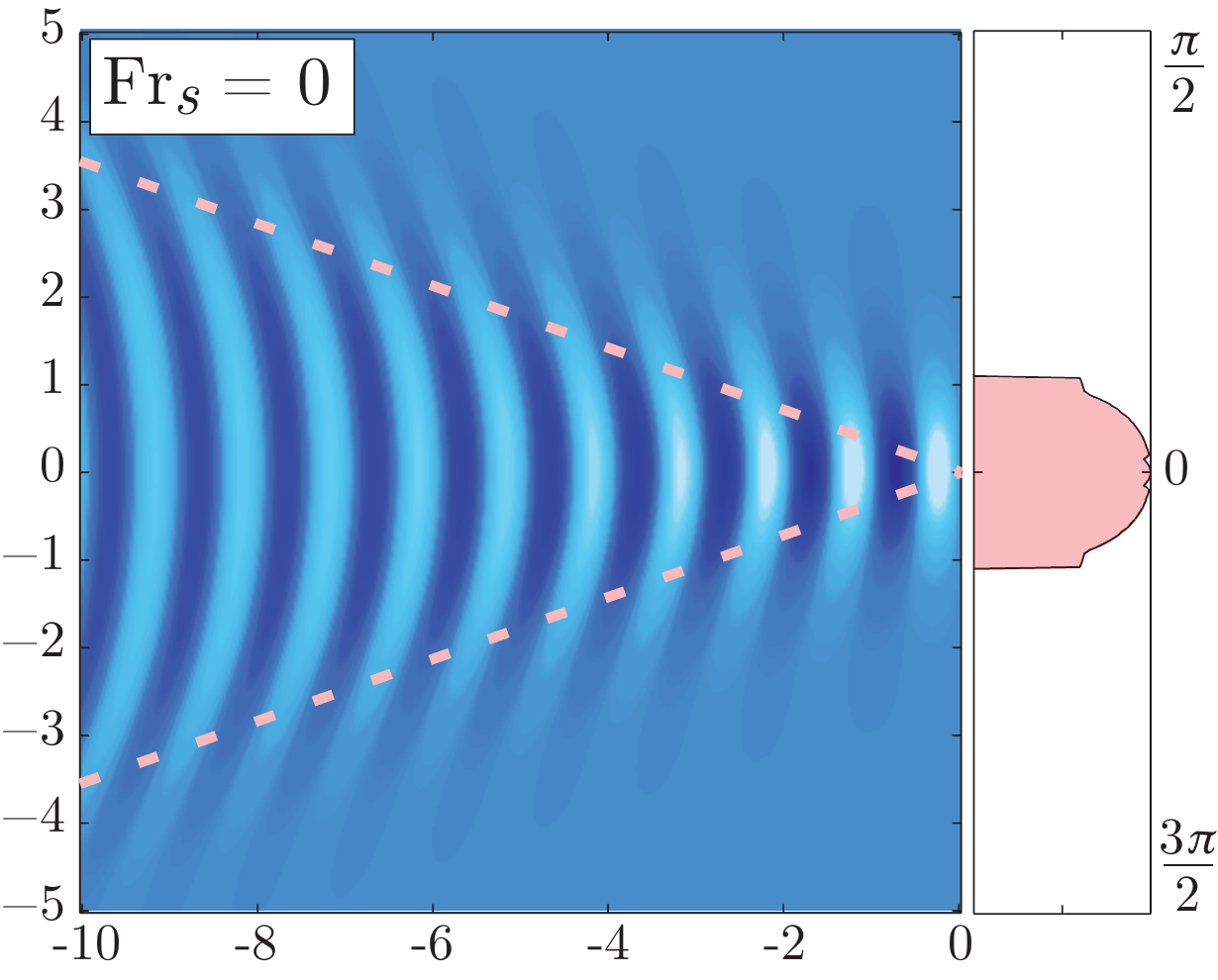}
  \includegraphics[width=.32\textwidth]{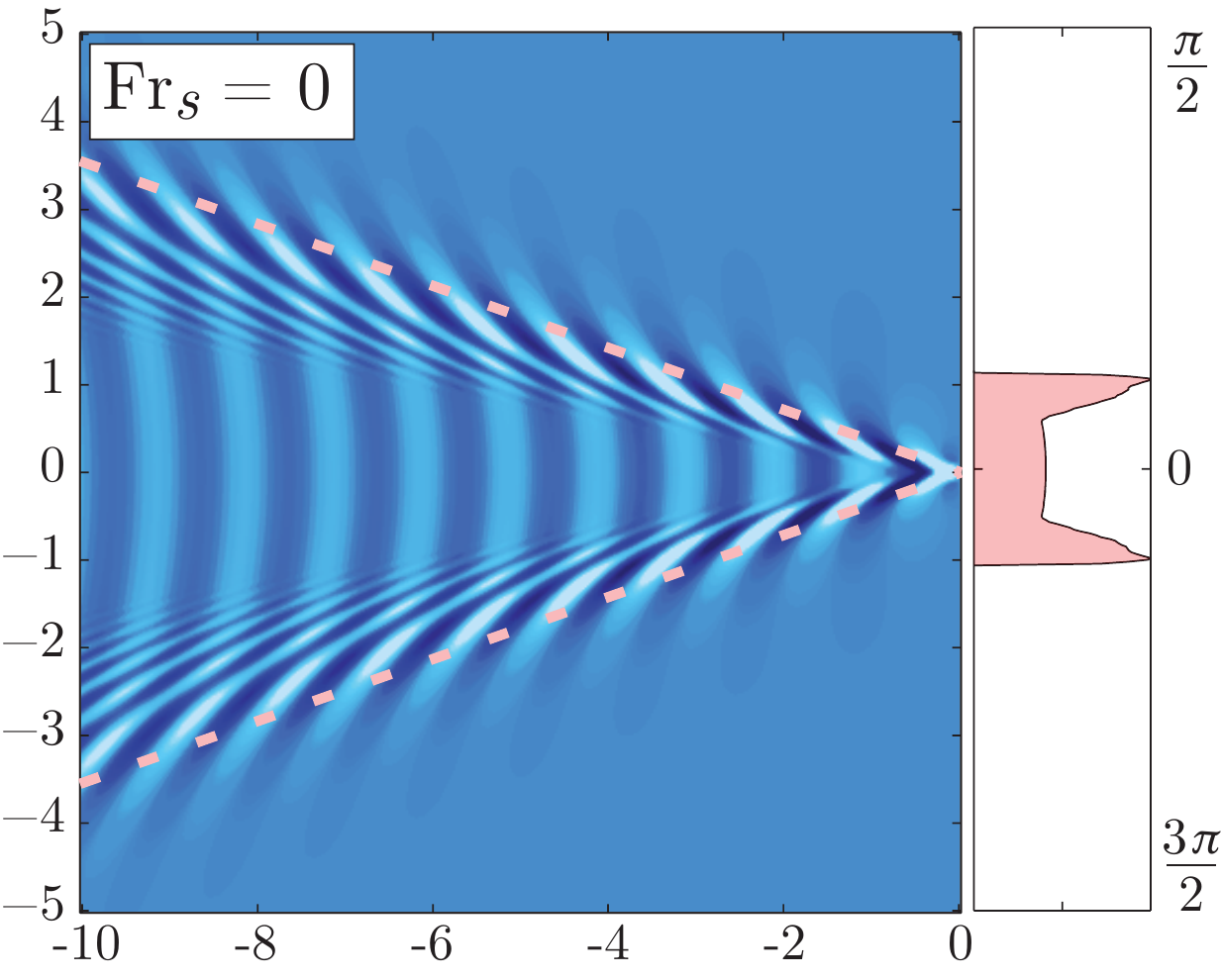}
  \includegraphics[width=.32\textwidth]{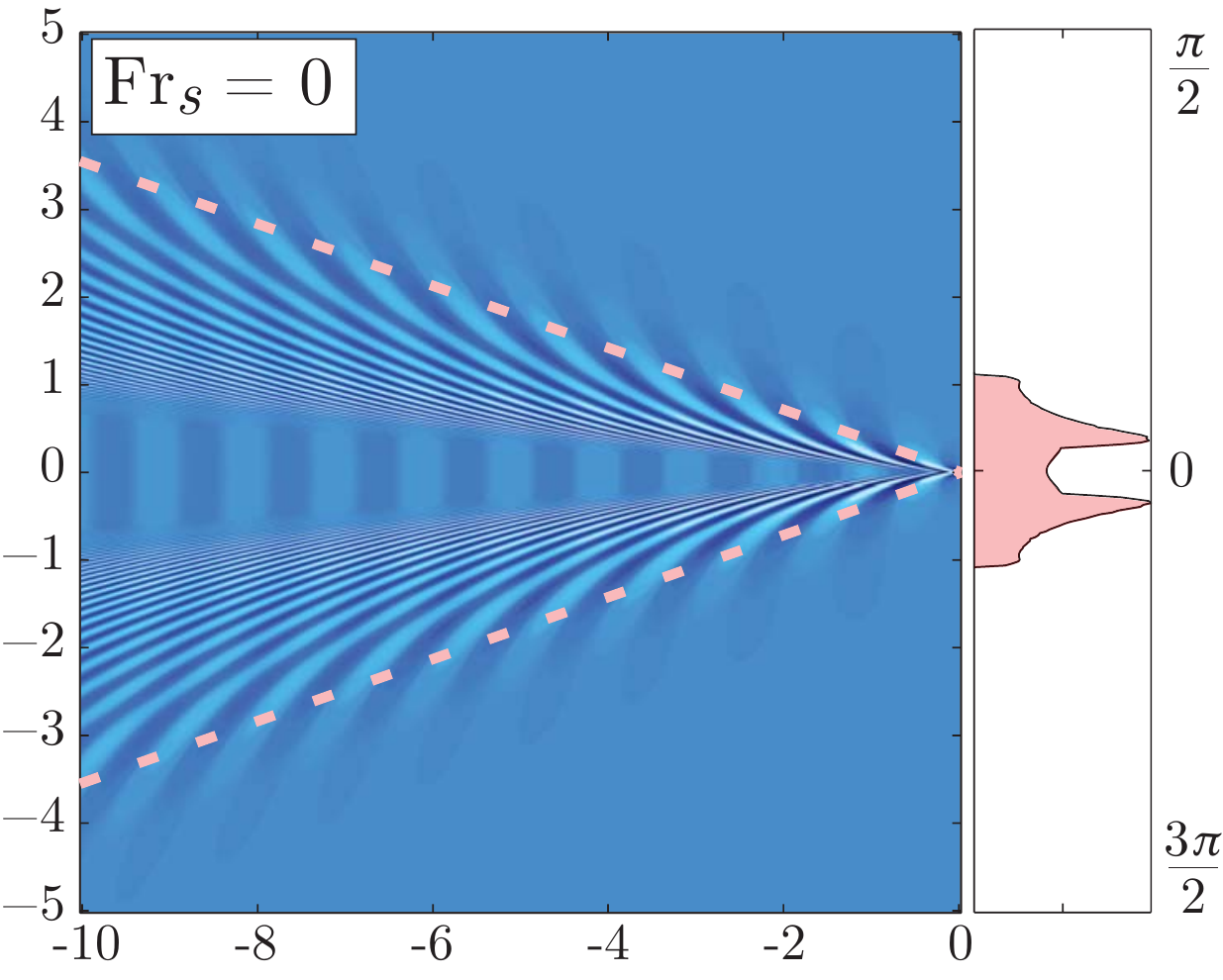}\\
  \includegraphics[width=.32\textwidth]{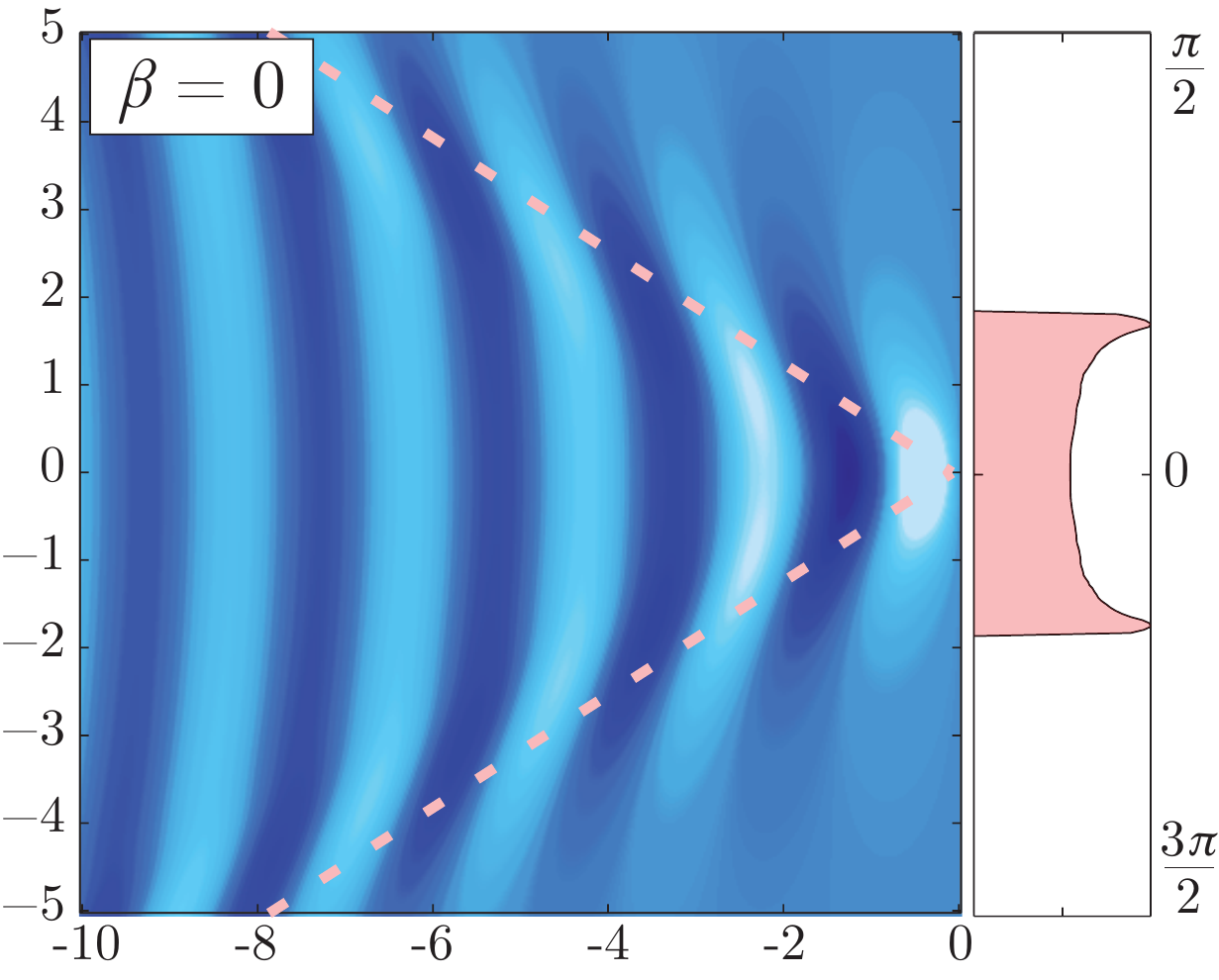}
  \includegraphics[width=.32\textwidth]{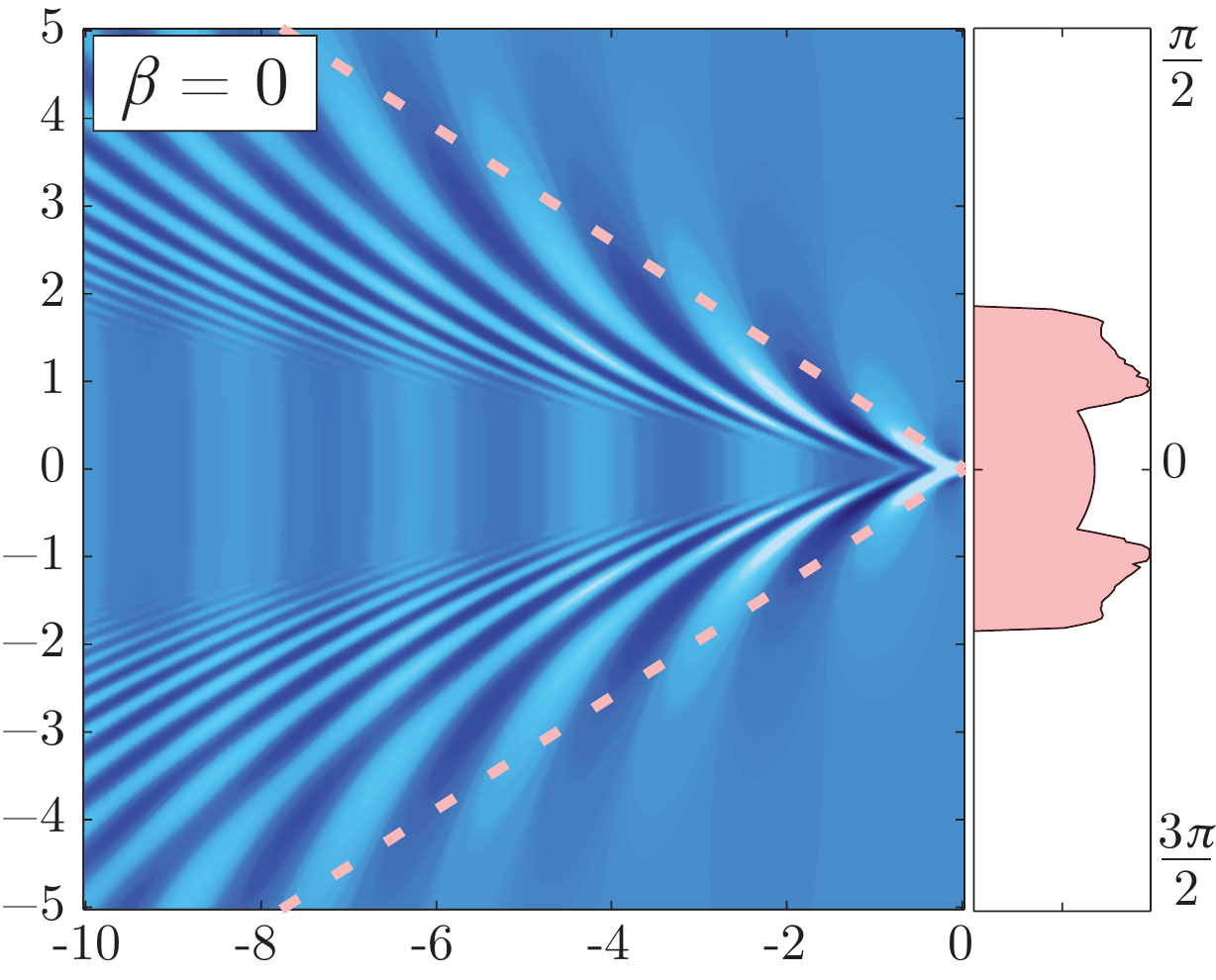}
  \includegraphics[width=.32\textwidth]{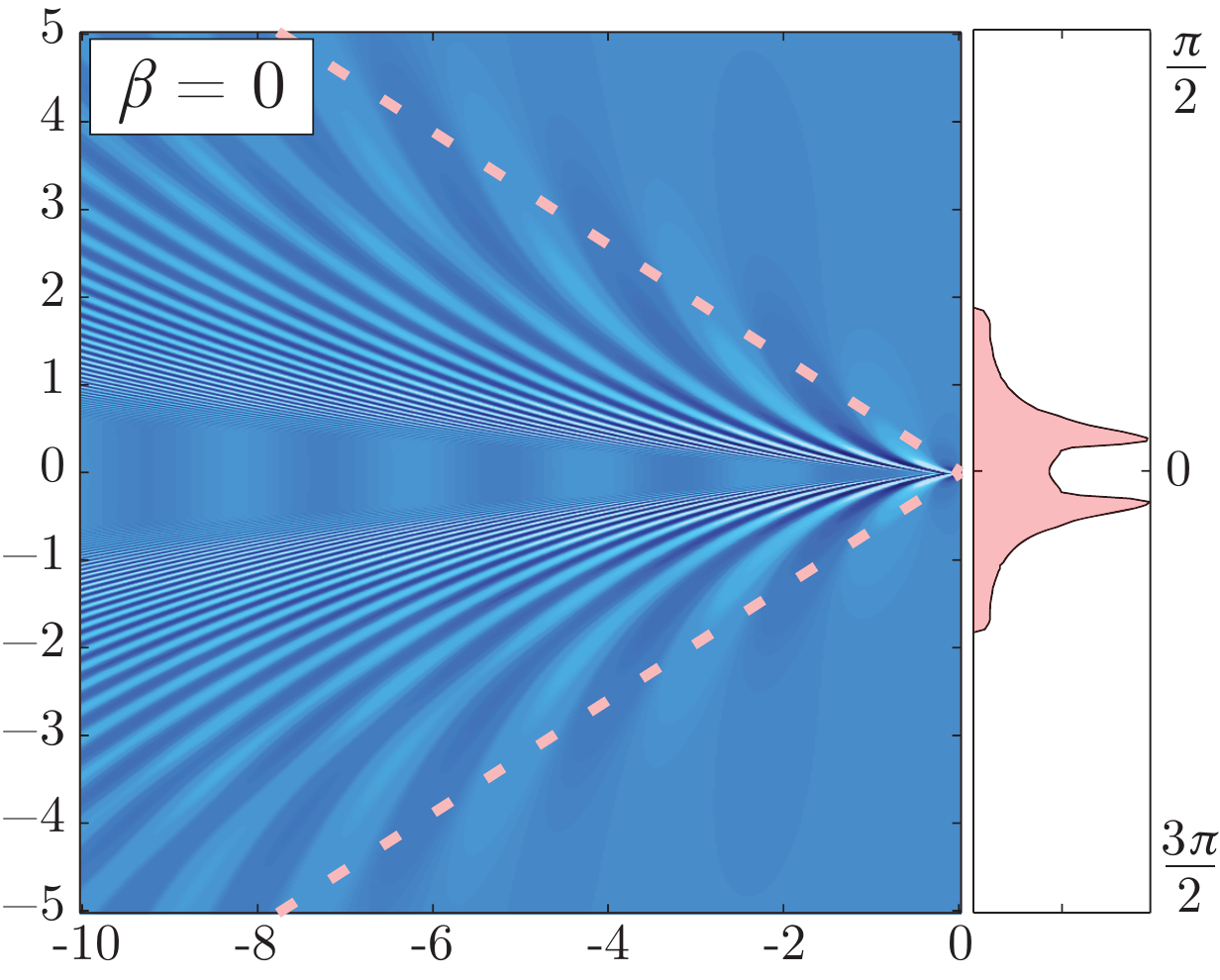}\\
  \includegraphics[width=.32\textwidth]{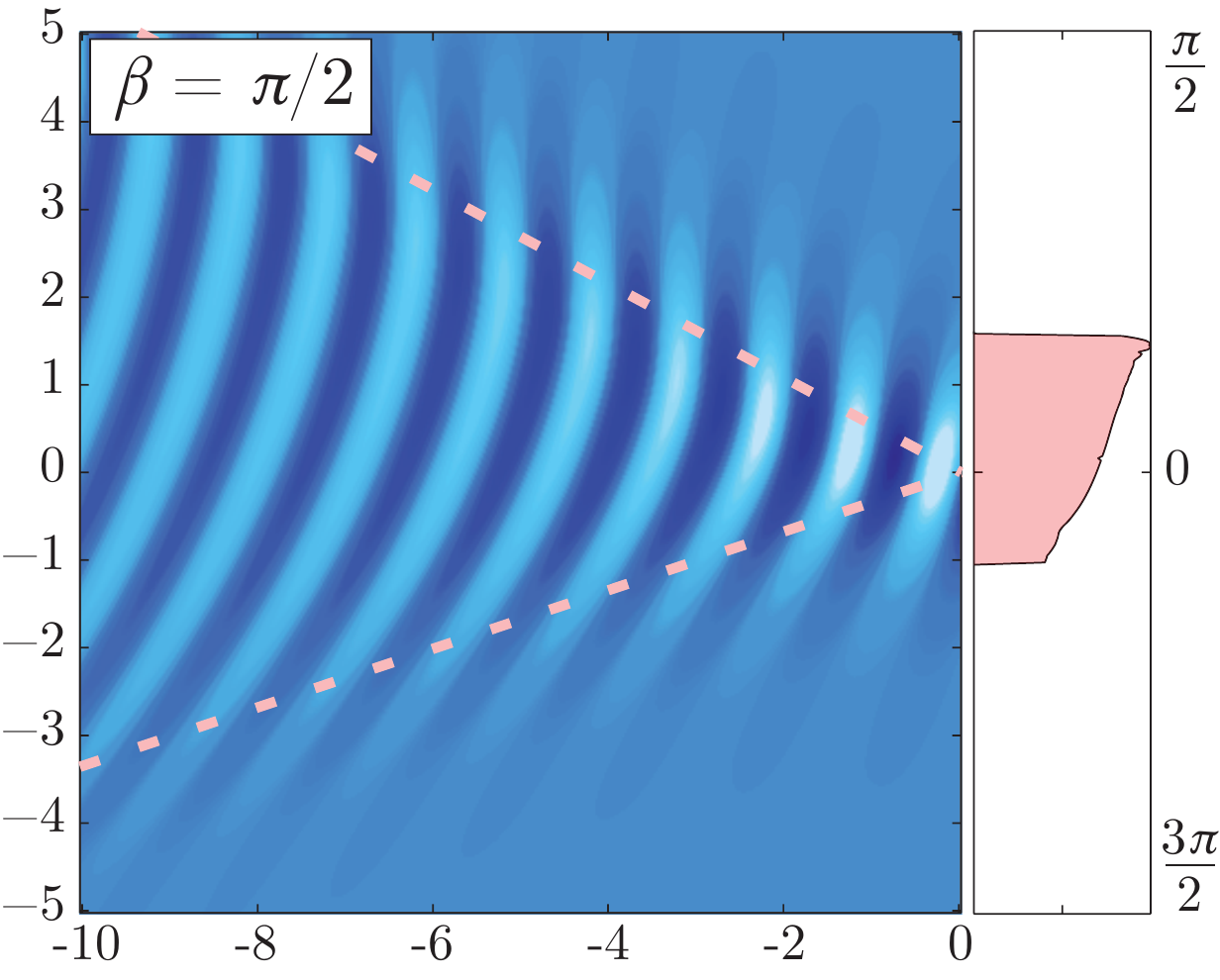}
  \includegraphics[width=.32\textwidth]{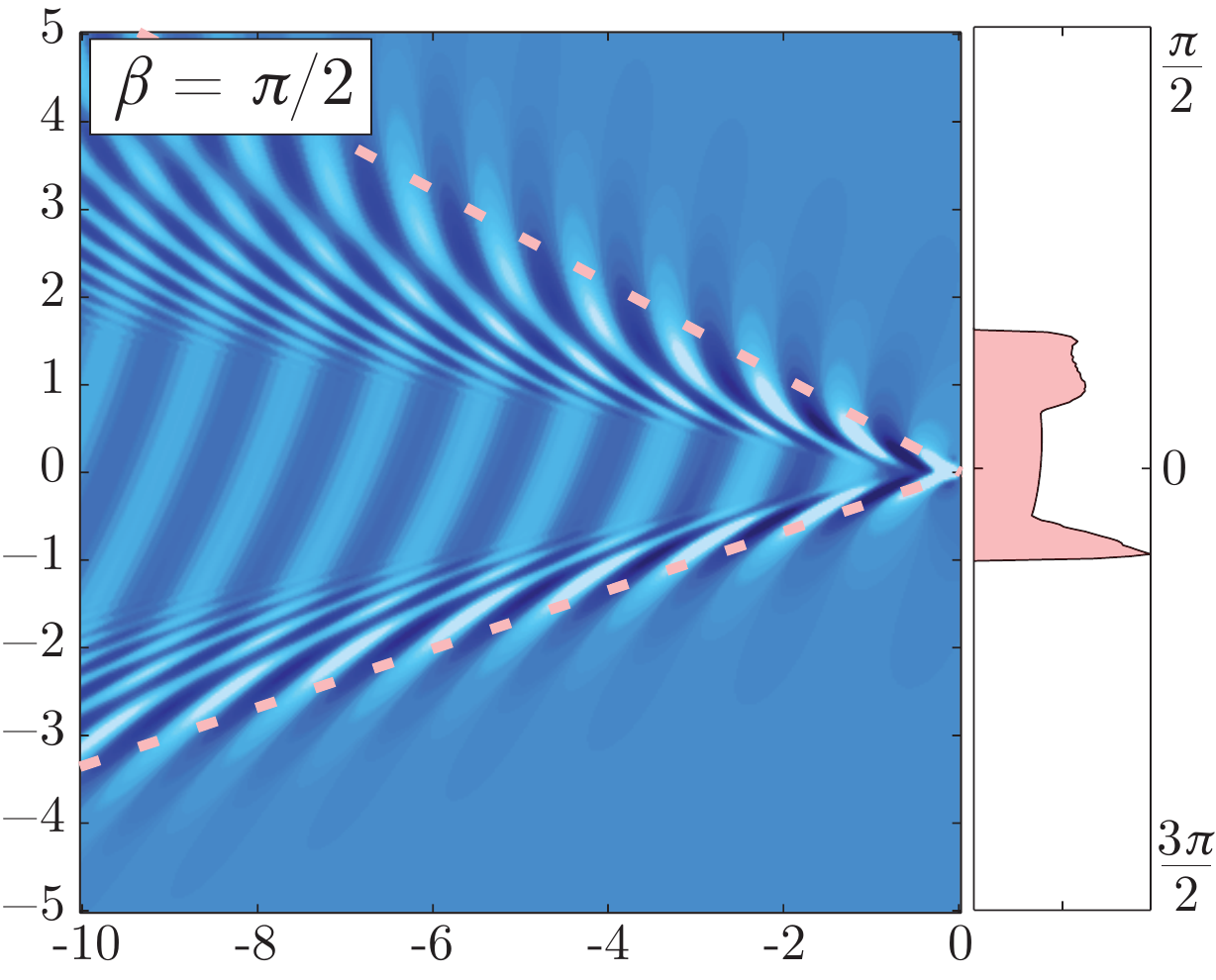}
  \includegraphics[width=.32\textwidth]{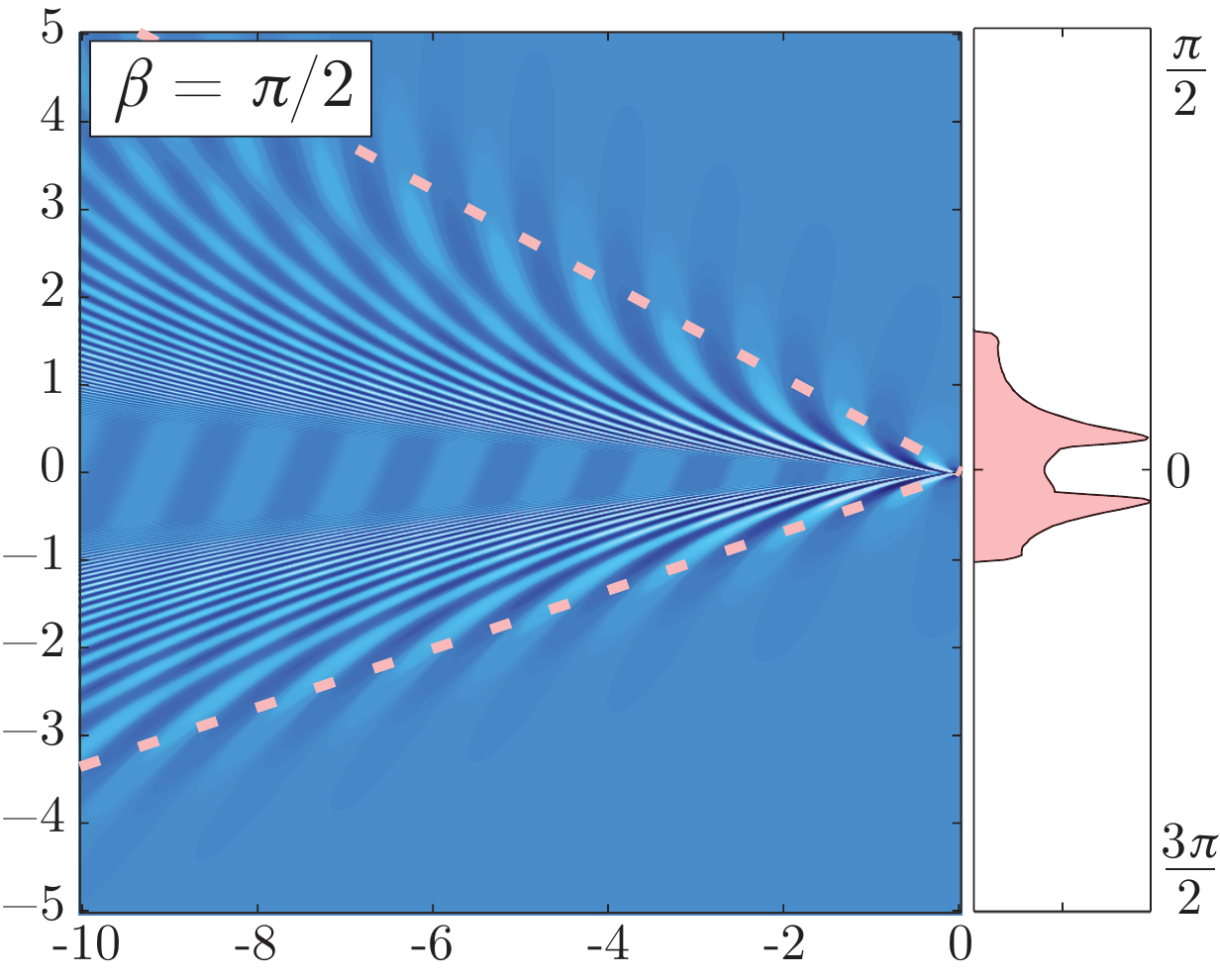}\\
  \includegraphics[width=.32\textwidth]{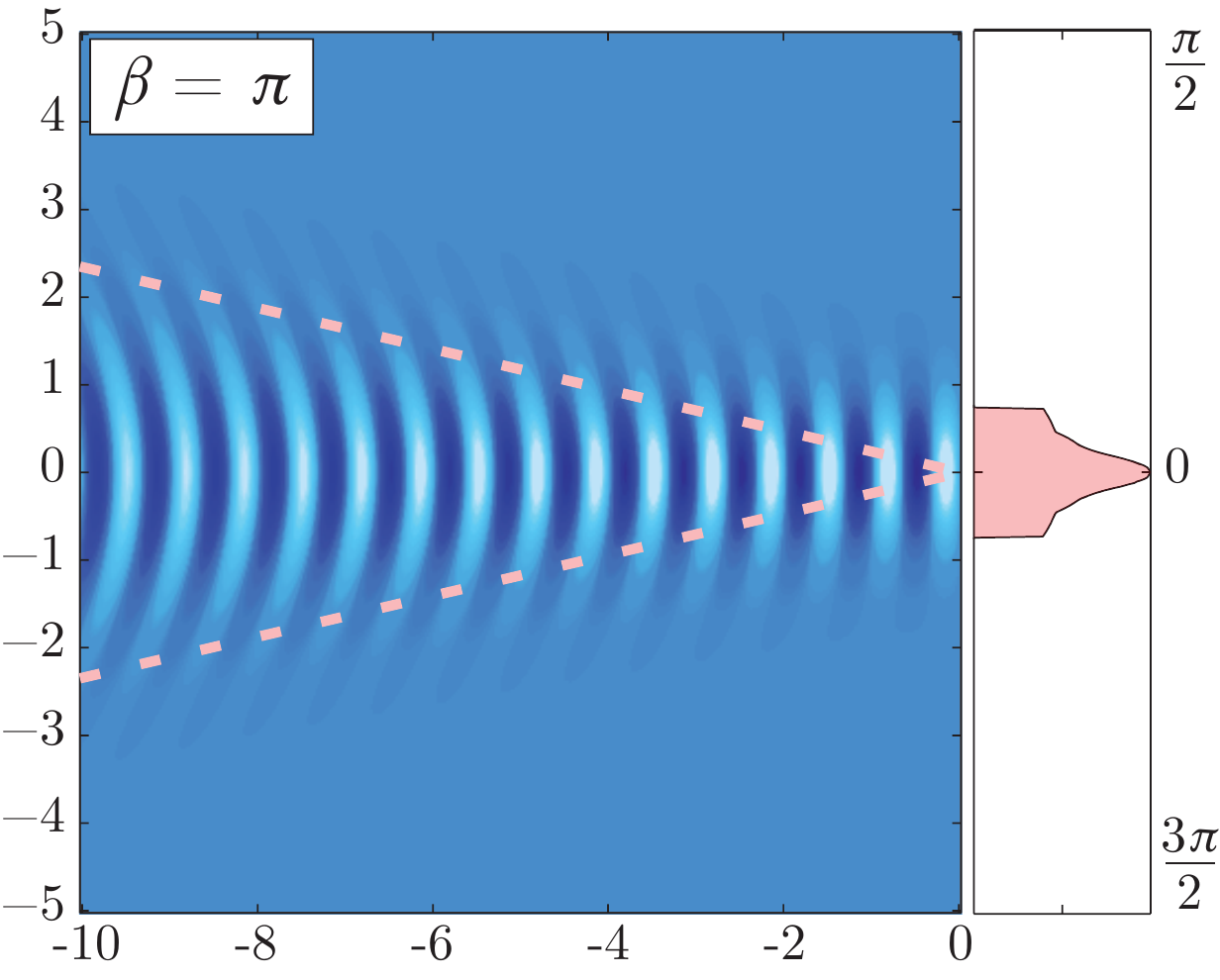}
  \includegraphics[width=.32\textwidth]{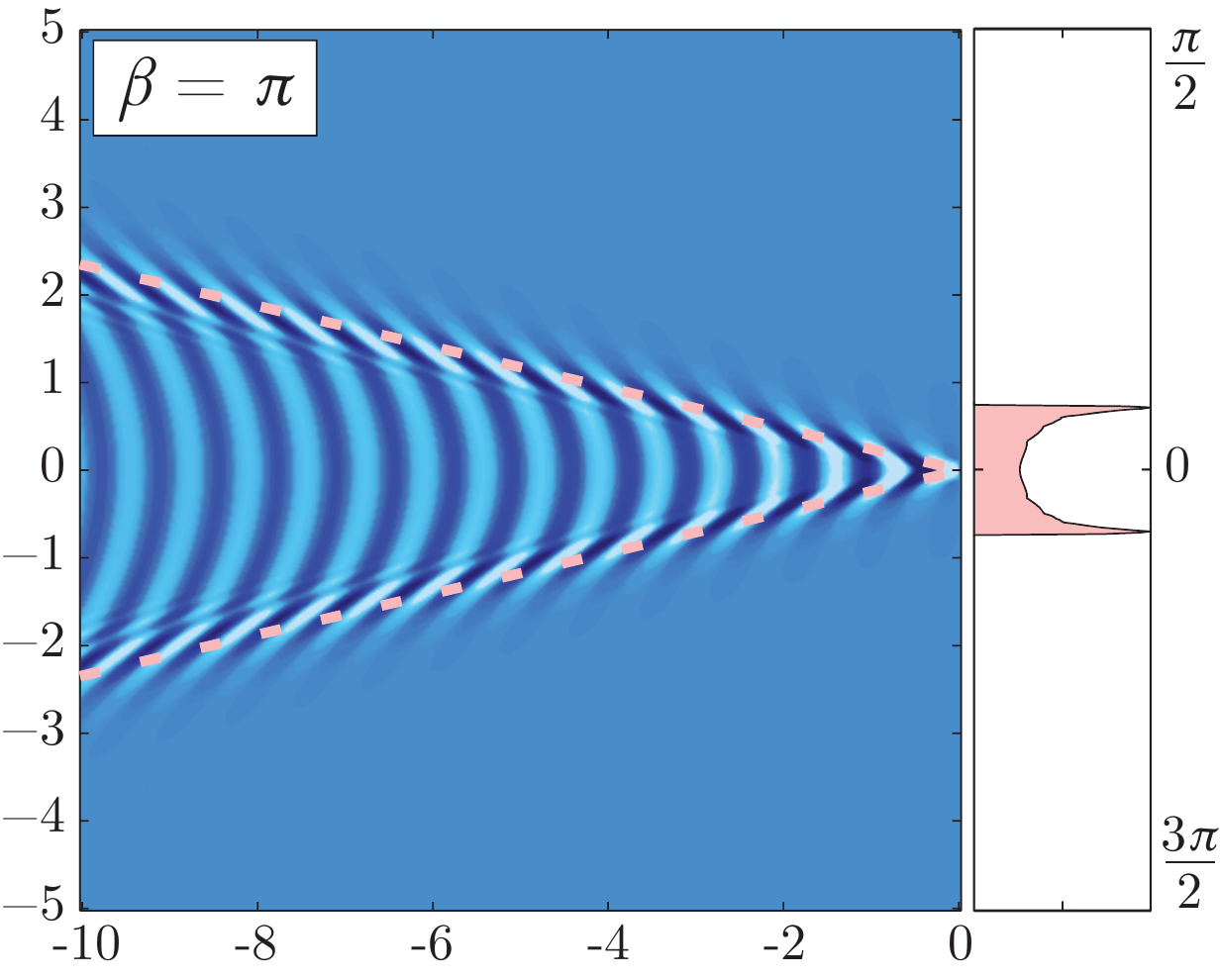}
  \includegraphics[width=.32\textwidth]{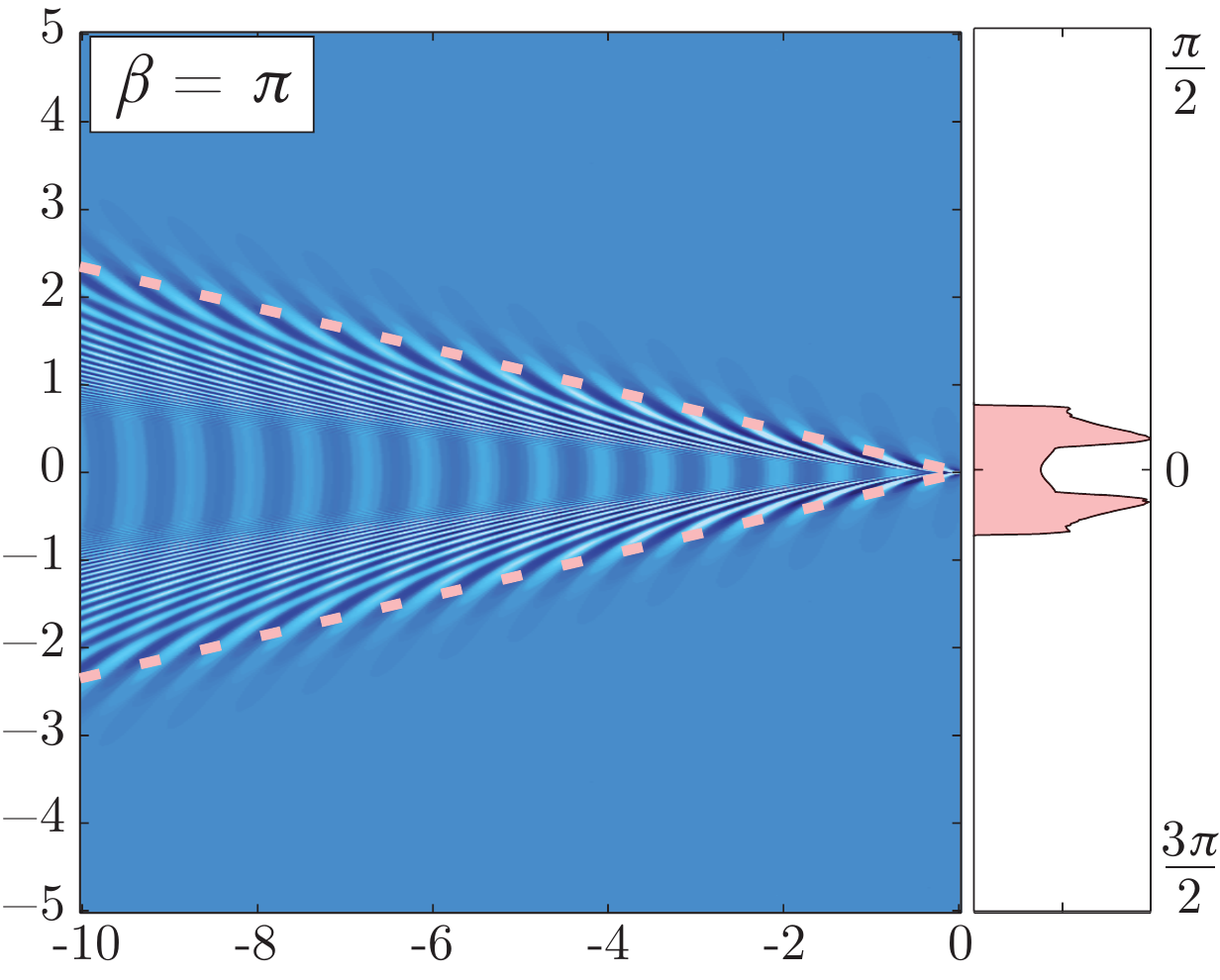}
  \end{center}
  \caption{Relief plots of wakes at $\Frs=0.5$, except top row at $\Frs=0$. Columns left to right show increasing source Froude numbers from slower/larger source (left) to faster/smaller source (right). Left to right: $\Fr=0.3, 0.8$ and $2.0$. Axes are in units of $\Lambda_0=2\upi b\Fr^2$. Relief shading (light to dark) has same scaling within each column. Panels to the right of each relief plot shows corresponding wave amplitude (rescaled by maximum value) as a function of angle $\phi-\beta$ between $\upi/2$ and $3\upi/2$. Kelvin angles are shown as dashed lines. A precise definition of the Kelvin angle and wave amplitude are given in section \ref{sec:num}.}
  \label{fig:plots}
\end{figure}

The term describing the propagating wake, proportional to $f(0)$ in Eq.~\eqref{sp}, gets contributions only where the denominator in Eq.~\eqref{B} is zero. Insisting $\mathbf{k}\cdot\mathbf{V}>0$ and noting $\mathbf{k}\cdot\mathbf{V}=kV\cos(\theta-\beta)$, one finds that this corresponds to
\be\label{steady}
  V\cos(\theta-\beta) = c(\mathbf{k}).
\ee
This is exactly the requirement for the wave pattern to be stationary as seen by the source, that the phase velocity along the direction of motion be equal to $V$, illustrated in the inset of Fig.~\ref{fig:phiK}.

Evaluating the far--term alone and letting $\gamma = \theta-\beta$, we find
\begin{align}
  \zeta({\boldsymbol\xi})\approx&-\frac{\zeta_0}{\Fr^4}\int_{-\frac{\upi}{2}}^{\frac{\upi}{2}}\rmd \gamma f_0(\gamma)\sin \Bigl[\frac{Rf_1(\gamma)}{\Fr^2}\Bigr]\rme^{-f_2^2(\gamma)},\notag \\  
  f_0(\gamma) =& \frac{f_s(\gamma)}{\cos^4\gamma};~~ 
  f_1(\gamma) = \frac{f_s(\gamma)\cos(\gamma+\beta-\phi)}{\cos^2\gamma},~~
  f_2(\gamma) = \frac{f_s(\gamma)}{2\upi \Fr^2\cos^2\gamma} \label{bigeq}\\
  f_s(\gamma)=& 1-\Frs\cos\gamma\cos(\gamma+\beta)\notag
\end{align}
with $R = r/b$, 
and the prefactor is $\zeta_0=p_0/(4\upi^2\rho g)$. The angle $\gamma$ as well as other angles used are defined in Fig.~\ref{fig:angles}.

\section{Numerical evaluation and results}\label{sec:num}

Eq.~\eqref{bigeq} was calculated numerically and the result is shown in Fig.~\ref{fig:plots}. Three different source Froude numbers are shown, $\Fr=0.3$, $0.8$ and $2$, to study the effect of a shear Froude number $\Frs=0.5$. The case of zero shear is shown for comparison (top row), and agrees perfectly with the results of \cite{darmon13}. Already at this modest exploration of the parameter space a wealth of interesting phenomena is manifest. For a source travelling against the shear ($\beta=0$) the wave pattern has larger Kelvin angle, and the transverse wavelength is longer. Both effects are opposite for motion along the shear direction ($\beta=\upi$), whereas the wake is asymmetric for side-on shear ($\beta=\upi/2$) with one side of the wake broader than the other, and the transverse waves directly behind the source propagating at an angle $\theta-\beta<0$.

Also the wave amplitude as a function of angle $\phi-\beta$ is plotted in panels to the right of each relief plot in Fig.~\ref{fig:plots}. By the method of stationary phase (see details in the following) one easily ascertains that $\zeta\sim 1/\sqrt{R}$ as $R\gg 1$ as required for equal measures of energy to pass through equal angular segments per unit time. Along a ray of constant $\phi$, $\sqrt{R}\zeta$ becomes a periodic function of $R$ with a constant maximum amplitude in a period, which is plotted, rescaled by the maximum value. 


\subsection{Critical shear Froude number} 

\begin{figure}
  \begin{center}
    \includegraphics[width=.32\columnwidth]{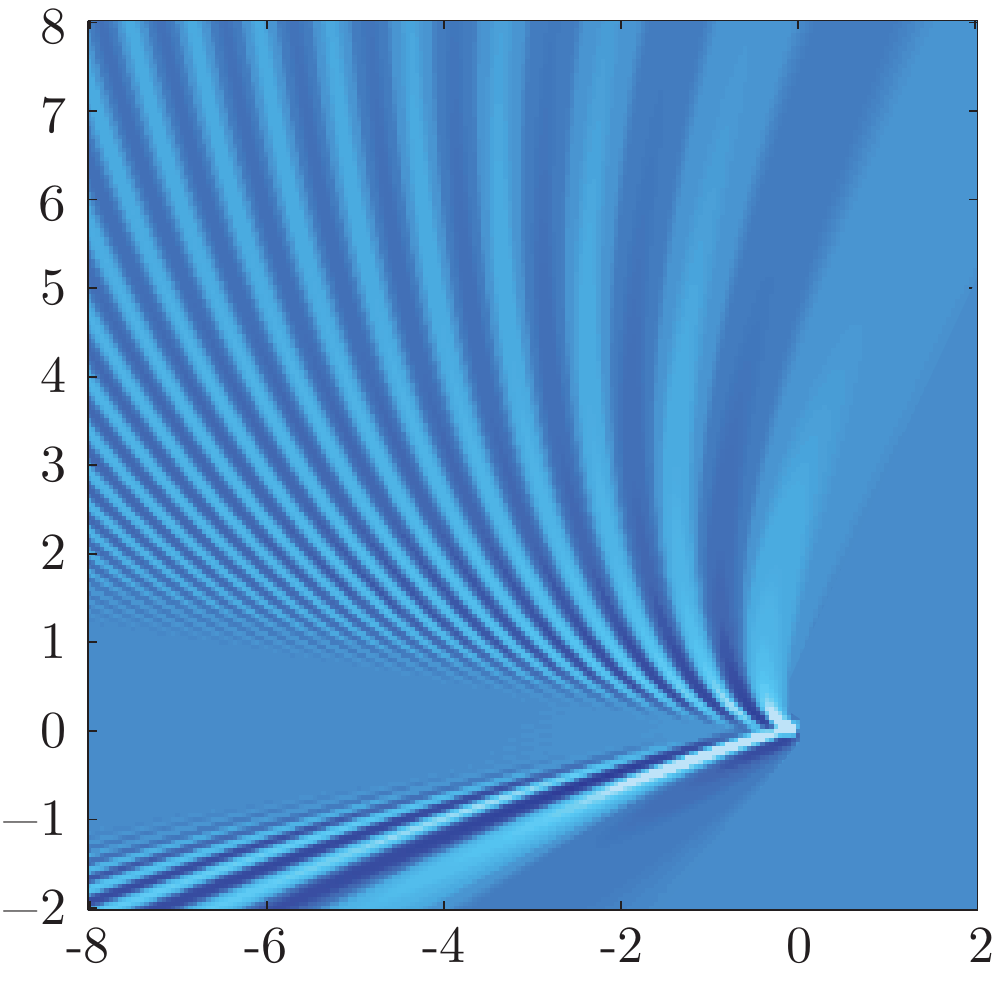}
    \includegraphics[width=.32\columnwidth]{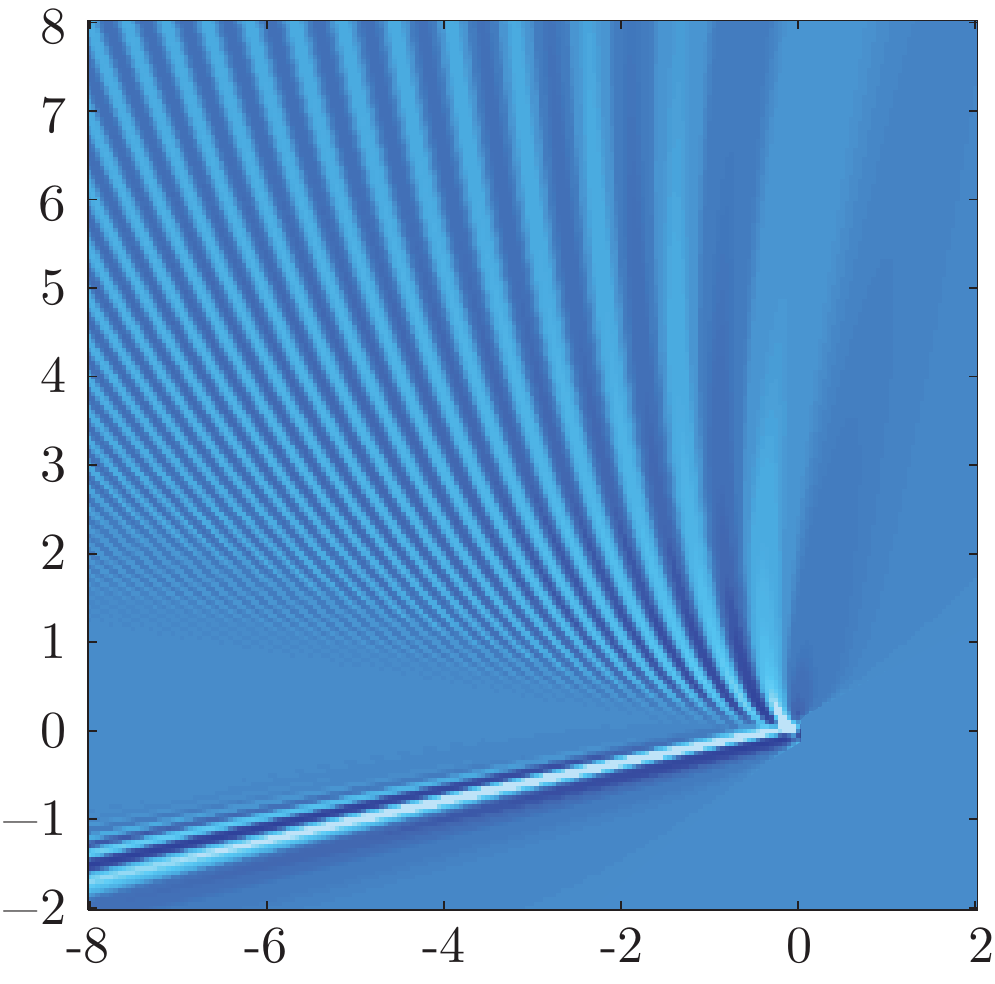}
    \includegraphics[width=.32\columnwidth]{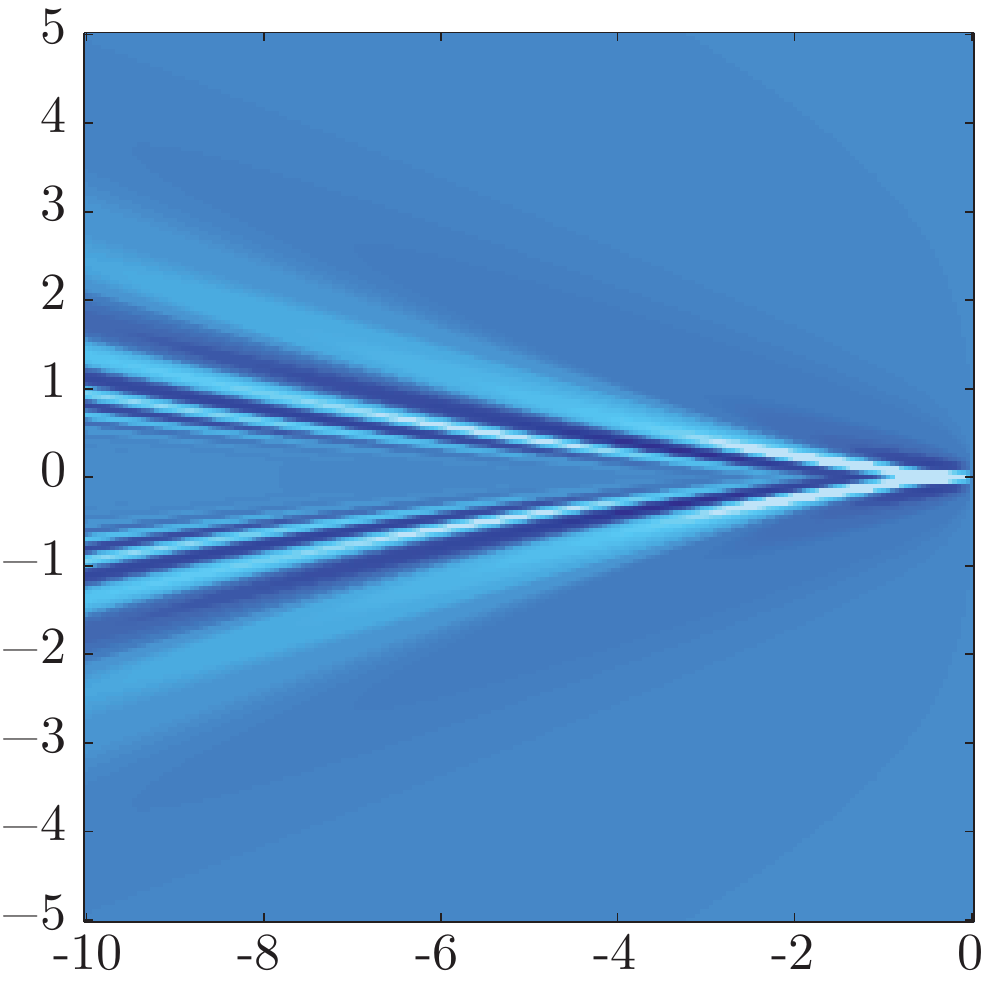}
  \end{center}
  \caption{Three examples of waves in extreme parameter regimes. Left: Close to the critical shear Froude number with side-on shear ($\Frs=1.99, \beta=\upi/2, \Fr = 0.8; \Fr_\mathrm{s,crit}=2$). Upper Kelvin angle exceeds 90$^\circ$. Centre: Supercritical velocity with side-on shear ($\Frs=4, \beta=\upi/2, \Fr = 0.8$). Upper Kelvin angle still exceeds 90$^\circ$. Right: Strongly supercritical velocity against the shear ($\Frs=5, \beta=0, \Fr=0.8; \Fr_\mathrm{s,crit}=1$). Axes in units of $\Lambda_0=2\upi b\Fr^2$.}
  \label{fig:3ex}
\end{figure}

Eq.~\eqref{deep} differs crucially from the dispersion relation in the absence of shear for long wavelengths, $k\to 0$. When $S=0$, the phase velocity scales as $\sqrt{g/k}$ and can be arbitrarily large for arbitrarily long waves. Here, the limit depends on the sign of $k_x=k\cos\theta$. The limiting behaviour when $k\to 0$ (or more precisely, $S|\cos\theta|\gg \sqrt{gk}$) is
\be\label{clim}
  c(k,\theta) \buildrel{k\to 0}\over{\longrightarrow} 
  \begin{cases}
    g/(S\cos\theta), &\text{for} \cos\theta>0 ,\\ 
    S|\cos\theta|/k,&\text{for} \cos\theta<0 .
  \end{cases}
\ee
At $\cos\theta=0$, $c(k)=\sqrt{g/k}$.
Hence, for waves whose direction of propagation has a component against the shear, $\cos\theta>0$, the phase velocity can never exceed $2g/S\cos\theta$. On the other hand, waves for which $\cos\theta<0$ increase in velocity with increasing wavelength as $k^{-1}$ instead of the usual $k^{-1/2}$, assisted by the shear flow.

An important consequence of this is that except for the direction $\mathbf{V}=(-V,0)$ there always exists a critical shear Froude number above which transverse waves vanish. 
The critical shear Froude number can be deduced from the stationary wave criterion, Eq.~\eqref{steady}, noting that when $\cos\theta >0$ the phase velocity $c(\mathbf{k})$ is bounded, Eq.~\eqref{clim}. Hence the stationary wave condition cannot be fulfilled in all propagation directions if the shear Froude number is so large that $V\cos(\theta-\beta)>c_\text{max}=g/S\cos\theta$, or in other words angles $\theta$ so that
\be\label{critagl}
  \cos(\theta-\beta)\cos(\theta) > \Frs^{-1}.
\ee
The integration range in Eq.~\eqref{bigeq} is thus restricted by
\be
  \gamma < -\frac1{2}[\arccos(2\Frs^{-1}-\cos\beta)+\beta] \text{ and } \gamma > \frac1{2}[\arccos(2\Frs^{-1}-\cos\beta)-\beta]
\ee
as well as $-\upi/2<\gamma<\upi/2$.
Angles $\gamma$ outside this range must be excluded from the integral in Eq.~\eqref{bigeq} since they cannot contribute to a stationary wake.
The maximum of the left hand side of Eq.~(\ref{critagl}) is at $\gamma = -\beta/2$ so the lowest value of $\Frs$ at which the waves in some propagation directions are too slow to keep up with the source is
\be\label{Frscrit}
  \Fr_\text{s,crit}=1/\cos^2(\beta/2).
\ee
We have shown some supercritical wave wakes in figure \ref{fig:3ex}.

\begin{figure}
  \begin{center}
    \includegraphics[width=.6\textwidth]{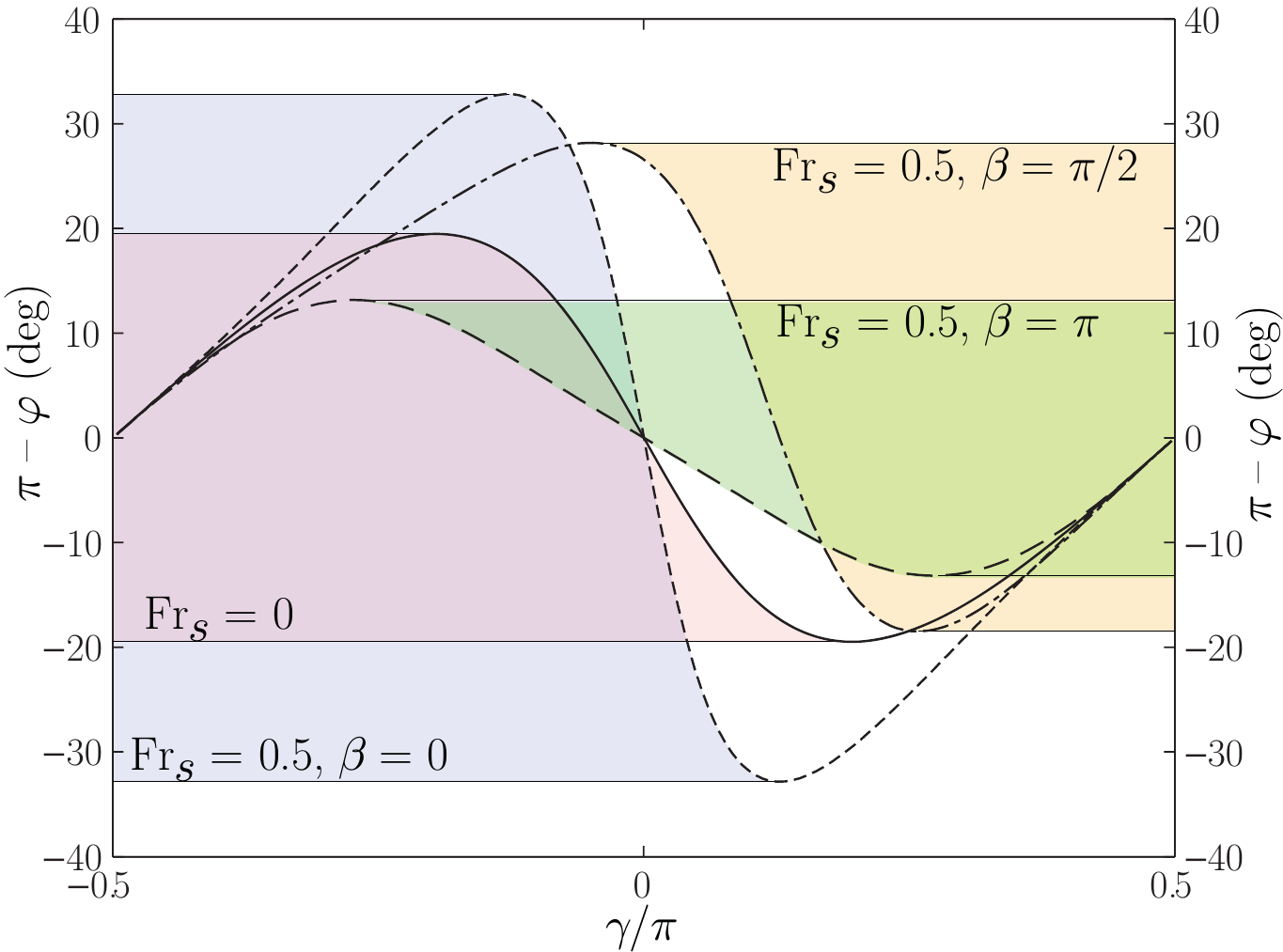}
  \end{center}
  \caption{Stationary point $\upi-\varphi(\gamma)$ [solutions of $f_1'(\gamma)=0$] for the parameter sets plotted in figure \ref{fig:plots}. Maxima and minima correspond to positive and negative Kelvin angles, shown as straight lines.}
  \label{fig:stat}
\end{figure}

As $\Frs\to\Fr_\text{s,crit}$, the wavelength of the transverse waves at $\phi-\beta=\upi$ increases \emph{ad infinitum}, and their group velocity 
\be
  c_g(k,\theta)=(\mathrm{d}/\mathrm{d}k)[k c(k,\theta)] = g/\sqrt{4gk + S^2\cos^2\theta}
\ee
approaches their phase velocity as $k\to 0$.

\subsection{Kelvin angle} \label{sec:KAgl}

The Kelvin angle $\phi_\mathrm{K}$ is found from integral \eqref{bigeq} using asymptotic analysis when $R\gg 1$. The integrand then contains a rapidly oscillating sine factor whose role is to limit the contributions to the integral to near its stationary points, i.e., where $f_1'(\gamma)=0$. Indeed, by the method of stationary phase, an integral of this form satisfies \citep[see][Ch.~6.5]{bender91})
\be\label{stph}
  \int_a^b\rmd x f(x)\rme^{\rmi R\phi(x)} \buildrel{R\to \infty}\over{\longrightarrow}
  \sqrt{\frac{2\upi}{R|\phi''(x_0)|}}if(x_0)\exp\Bigl\{\rmi R\phi(x_0)+\rmi\frac{\upi}{4}\mathrm{sg}[\phi''(x_0)]  \Bigr\}
\ee
provided that there is a point $x_0$ where $\phi'(x_0)=0$ and $\phi''(x_0)\neq 0$, and  $a\leq x_0\leq b$. The integral behaves asymptotically as $1/\sqrt{R}$ as previously promised. 

Stationary phase equation $f_1'(\gamma)=0$ amounts to picking out the propagation directions $\gamma$ which makes significant contributions at a particular angle $\phi$. Letting $\varphi=\phi-\beta$ (see Fig.~\ref{fig:angles}), we plot the stationary phase solution $\upi-\varphi(\gamma)$ as function of $\gamma$ in Fig.~\ref{fig:stat}, for the parameters used in Fig.~\ref{fig:plots}.

The Kelvin angle is the largest value of $\upi-\varphi$ for which a stationary point $f_1'(\gamma)=0$ exists. It is thus independent of the source Froude number $\Fr$ \citep{darmon13}, but depends crucially on the shear number $\Frs$. At angles $\phi$ lying strictly inside the Kelvin wedge, two propagation directions $\gamma$ contribute. The one closest to zero is of transverse type, while that closest to $\pm \upi/2$ is of diverging type. 

\begin{figure}
  \begin{center}
    \includegraphics[width=.65\columnwidth]{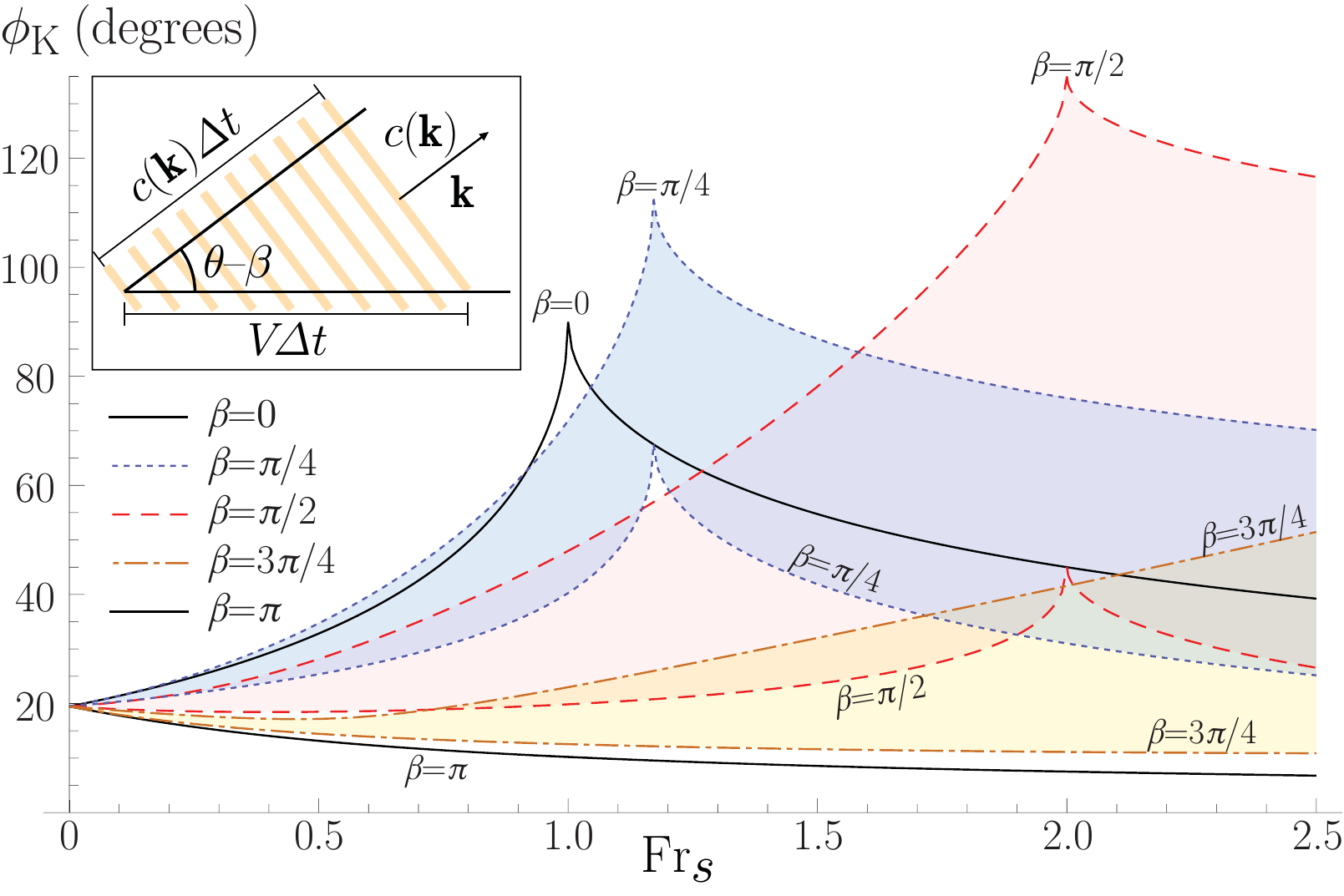}
  \end{center}
  \caption{Kelvin angles as function of $\Frs$ for different $\beta$. Shaded areas connect graphs pertaining to the same asymmetric wake. Inset: Illustration of steady wave relation Eq.~\eqref{steady}.}
  \label{fig:phiK}
\end{figure}

We have calculated the Kelvin angle $\phi_\mathrm{K}$ numerically at different $\Frs$ and $\beta$, as plotted in Fig.~\ref{fig:phiK}. 
As the critical shear Froude number is approached, the sum of the Kelvin angles on either side of the wake combine to a total wake angle of $\upi$. This behaviour is closely similar to that encountered in shallow water, see \citet[see][\S 15]{havelock08}, where the phase velocity is also bounded in the long wavelength limit, albeit for a different reason. For a side-on shear current, the Kelvin angle on one side will then exceed $\upi/2$. Kelvin angles $\phi_\mathrm{K}$ are plotted as function of $\Frs$ for different $\beta$ in figure \ref{fig:phiK}. For supercritical $\Frs$, Kelvin angles are monotonously decreasing functions of $\Frs$. 

\subsection{``Mach angle''} 

The vertical panels of Fig~\ref{fig:plots} for $\Fr=2$ show how the wave amplitude has a clear peak at a certain angle $\phi-\beta$ for large Froude numbers. It was shown by \cite{darmon13} and \cite{rabaud13} that this angle, $\phi_\mathrm{M}$, which we refer to as the ``Mach'' angle in analogy with a shock wave, scales as $\mathrm{Fr}^{-1}$ for large $\Fr$ in the absence of shear. 

The effect of $\Frs$ on the scaling of $\phi_\mathrm{M}$ can be found by considering the stationary phase approxiation of Eq.~\eqref{bigeq}, 
\begin{align}
  \zeta \approx -\frac{\sqrt{2\upi}\zeta_0}{\Fr^3\sqrt{R}}\sum_{k={\mathrm{t},\mathrm{d}}}\frac{f_s(\gamma_2)\exp[-f_2^2(\gamma_k)]}{\cos^4\gamma_k\sqrt{|f_1''(\gamma_k)|}}\cos\Bigl\{\frac{R f_1(\gamma_k)}{\Fr^2}+\frac{\upi}{4}\mathrm{sg}[f_1''(\gamma_k)]\Bigr\}\label{asym}
\end{align}
where $\gamma_\mathrm{t}$ ($\gamma_\mathrm{d}$) are the propagation directions of transverse (diverging) wavefronts, and calculating the maximum of the amplitude with respect to $\phi$ under the assumption $\Fr\gg 1$, following essentially the same procedure as \cite{darmon13}. For large $\Fr$ one finds that the transverse waves are negligible \citep{darmon13}, and by solving for the maximum of the amplitude in Eq.~\eqref{asym} we obtain
\be
  \phi_\mathrm{M}\sim \frac{1}{40^{1/4}\sqrt{\upi}}\frac1{\Fr} + \frac{3\sqrt{10}}{80\upi}\frac{\Frs\sin\beta}{\Fr^2}+....
\ee
The first term is exactly that found by \cite{darmon13}. While the constant coefficients depend on the exact form of $p_\text{ext}$ and are not so significant, the scaling is general and shows that the ``Mach'' angle is hardly influenced by the shear current for large $\Fr$.

\subsection{Further discussion} 
Inspecting Fig.~\ref{fig:plots}, the wavelength of transverse waves is lengthened for $\beta=0$ and shortened for $\beta=\upi$. Solving for stationary points at $\phi-\beta=\upi$ assuming $\Frs<\Fr_\text{s,crit}$ easily yields the wavelength of transverse waves in these two cases:
\be
  \Lambda(\beta=0)=\frac{\Lambda_0}{1-\Frs}; ~~ \Lambda(\beta=\upi)=\frac{\Lambda_0}{1+\Frs}; ~~ \Lambda_0=2\upi b\Fr^2
\ee
where the $\beta=0$ case assumes $\Frs<\Fr_{\mathrm{s},\text{crit}}=1$. 
For a general $\beta$ the transverse waves at $\phi-\beta=\upi$ propagate at an angle $\gamma_t$ which solves
\be
  \tan\gamma_t\sec\gamma_t = -\Frs \sin(\gamma_t+\beta).
\ee
When $\Frs=\Fr_{\mathrm{s, crit}}$ from Eq.~\eqref{Frscrit}, the solution is $\gamma_t=-\beta/2$, which we found was exactly the angle at which the maximum velocity is first reached. In other words, the critical Froude number is the point at which the transverse waves directly behind the source obtain a group velocity which equals the phase velocity, hence cannot propagate energy away.

\section{Concluding remarks} 
We have solved the linear ship wave problem in the presence of a shear flow of uniform vorticity. The Kelvin angle limiting the sector within which ship waves can propagate is no longer constant at deep water, but depends significantly on the strength and direction of the vorticity $S$, as characterized by a ``shear Froude number'' $\Frs=VS/g$ ($V$ is the ship's velocity, $S$ the constant vorticity) and an angle $\beta$. Except when the ship motion is exactly that of the shear flow there exists a critical velocity $V$ at which the combined Kelvin angle reaches $180^\circ$, and above which no transverse waves exist, similar to the situation in shallow water. With side-on shear, the Kelvin angle on one side of the wake will exceed $90^\circ$ close to the critical velocity. The Kelvin angles grow as a function of velocity up to the critical value, whence it decreases, similar to what is found for ship waves in shallow water.

The angle where wave amplitude is maximal was shown by \cite{rabaud13} and \cite{darmon13} to scale as the inverse of $\Fr$, the Froude number based on the ship size, when $\Fr\gg 1$. This scaling is largely unaffected by the presence of shear, at sub--leading order in $\Fr^{-1}$. The solution presented in this Communication is readily extended to finite depth and non--negligible surface tension, and displays a rich set of physical phenomena for future investigation. We have benefited from discussions with I.~Brevik and P.~Tyvand in the preparation of this Communication.

\bibliographystyle{jfm}
\bibliography{shipwave}

\end{document}